\documentclass[aps,twocolumn,showpacs]{revtex4}
\usepackage{graphics}
\begin{document}
\tighten
\bibliographystyle{apsrev}
\def\half{{1\over 2}}
\def \D {\mbox{D}}
\def\curl {\mbox{curl}\,}
\def \ep {\varepsilon}
\def \lleq {\lower0.9ex\hbox{ $\buildrel < \over \sim$} ~}
\def \ggeq {\lower0.9ex\hbox{ $\buildrel > \over \sim$} ~}
\def\beq{\begin{equation}}
\def\eeq{\end{equation}}
\def\ber{\begin{eqnarray}}
\def\eer{\end{eqnarray}}
\def \apl {ApJ, }
\def \aps {ApJS, }
\def \pd {Phys. Rev. D, }
\def \prl {Phys. Rev. Lett., }
\def \pl {Phys. Lett., }
\def \np {Nucl. Phys., }
\def \l {\Lambda}

\title{A NOTE ON THE COSMOLOGICAL DYNAMICS \\
IN  FINITE-RANGE GRAVITY}
\author{M. Sami}
\email{sami@iucaa.ernet.in}
\affiliation{Inter-University Centre for Astronomy and Astrophysics,
Post Bag 4, Ganeshkhind, Pune-411 007, INDIA.\footnote{On leave from jamia Millia, New Delhi.}}


 \pacs{98.80.Cq,~98.80.Hw,~04.50.+h}

 \begin{abstract}
 In this note we consider the homogeneous and isotropic cosmology in the finite-range gravity theory recently proposed by
 Babak and  Grishchuk. In this scenario the universe undergoes late time accelerated expansion if both the massive gravitons present in the model are  tachyons. We carry out the phase space analysis of the system and show that the late-time acceleration is an attractor of the model.

 \end{abstract}
  \maketitle
 \section{INTRODUCTION}
 Recently there has been renewed interest in the old idea that gravitons could have a small mass \cite{pauli}. The fresh motivation for massive gravity may be attributed to the high precision experiments on the detection of gravitational waves like LIGO and LISA which, in a near
 future, may result in answering the question about the mass of gravitons\cite{will,larson,
cutler}. Secondly it seems to be natural to look for an 
 alternative to GR to  incorporate the recently observed "late time accelerated expansion" of the universe. Third, the higher dimensional gravity theories seem to naturally mimic the properties of four dimensional massive gravity and should be given serious thought 
\cite{dvali}. Last but not the least it is more than desirable to incorporate the possibility of inflation without opting for an ad hoc mechanism to realize it. The massive gravity may or may not address all these issues.\par 
 Till very recently, the masses of gravitons were thought to come from Fierz-Pauli mass term added to Einstein Hilbert action.
The model faces the well known mass discontinuity problem called Van Dam-Veltman-Zakharov 
discontinuity\cite{iwasaki,vandam,zakharov,vains,Fadeev,visser,kogan,porati,gulini,dvali,desser,gruzinov,veltman}.It was very recently
noticed by Babak and Grishchuk that this problem is very specific to the choice of Fierz-Pauli mass term\cite{babak1}.
In field theoretic formulation of Babak and Grishchuk\cite{babak1,babak}, the mass term has the form, $ k_1 h^{\mu \nu} h_{\mu \nu}+k_2h$ , where $ h^{\mu \nu}$ is the nonlinear gravitational field propagating in the Minkowski space and "h" denotes the tress of $ h^{\mu \nu}$. Linear massive gravity is discussed in Refs.\cite{barinov,mahesh,lee}.
 For a specific choice of parameters $k_1$ and $k_2$: $ k_1+k_2=0$, the local predictions of the model have finite difference with those of GR though the mass term in the Lagrangian smoothly vanishes for $k_1 \rightarrow 0$. Babak and Grishchuk have demonstrated that everywhere  on the parameter plane $(k_1,k_2)$ excluding the
  Fierz- Pauli straight line( $k_2=-k_1$ ), the massive gravity theory proposed by them is free from discontinuity problem\cite{babak1}. The parameters
 $k_1$ and $k_2$ are related to the masses of gravitons: $\alpha$ - mass of spin-2 graviton and $\beta$ - mass of the spin-0 graviton. The predictions of Babak-Grishchuk model are dramatic: their model has smooth limit to GR for $\alpha, \beta  \rightarrow 0$, the inclusion of a small mass
 term removes the black hole event horizon and makes the cosmological evolution oscillatory. Further, it is  remarkable that if both the
 gravitons in the model are sacrificed to tachyons, the finite range gravity exhibits late time accelerated expansion of the universe. The miracle is done by the presence of massive scalar graviton. In case, the mass of scalar graviton is zero, Babak and Grishchuk have noticed that cosmology based upon the finite range gravity is identical to GR cosmology independently of the mass of spin-2 graviton\cite{babak1}. Hence the additional spin-0 graviton plays a central roll in  finite-range gravity. For earlier references on field theoretic formulation of gravity see Refs\cite{feynman,petrova,zel} and the work of Logunov and collaborators\cite{logunov}. The non-linear bigravity theory recently discussed by Damour et al. seems to share many interesting features of the finite range gravity mentioned above \cite{damour}.

Once the Fierz-Pauli choice for mass parameters is excluded, the success
of the theory relies on existence of an additional scalar graviton. This is 
reminiscent of the manner in which a massive scalar boson was introduced
by Veltman in the electroweak theory \cite{veltman}, and suggestion
contained in Ref. \cite{dva} that gauge bosons and  massive graviton
must necessarily be interpreted as multi-spin objects. In particular,
it is anticipated that a finite-range gravity should endow graviton
with spin 0, 1 and 2  components. However, in this work we shall
strictly confine to the framework   of Babak and Grishchuk.   

\par
 In this note we investigate the phase space behavior of cosmological evolution in finite range gravity with both the gravitons behaving like 
 tachyons. We demonstrate that the accelerated expansion is a late time 
attractor of the model.

 \subsection{EVOLUTION EQUATIONS IN FINITE RANGE GRAVITY}
 The gravitational field in the massive gravity is described by a non-linear tensor field $ h^{\mu \nu}$ propagating in the Minkowski
 space with metric $\gamma^{\mu \nu}$ and with a massive term added to the GR part of the field Lagrangian which is quadratic in $h^{\mu \nu}_{;\tau}$ (covariant derivative is taken with respect to the metric $\gamma^{\mu \nu}$). The
effective field equations have the form\cite{babak1}
\begin{equation}
 G_{\mu \nu}+ M_{\mu \nu}= T_{\mu \nu}
\end{equation}
 where $T_{\mu \nu}$ is the matter energy momentum tensor and the mass tensor $ M_{ \mu \nu}$ is given by
\begin{equation}
 M_{ \mu \nu}=\left(\delta_{\mu}^{\alpha} \delta_{\nu}^{\beta}-{1\over 2} g^{\alpha \beta} g_{\mu \nu} \right) \left(2k_1h_{\alpha \beta}
 +2 k_2\gamma_{\alpha \beta }h \right)
\end{equation}
 where $\sqrt{-g} g^{\mu \nu}=\sqrt{-\gamma}\left(\gamma^{\mu \nu}+ h^{\mu \nu}\right)$. In a homogeneous and isotropic situation
 $h_{\mu \nu}$ depends upon time alone and the effective metric acquires the form
 \begin{equation}
 ds^2=b^2(t)dt^2-a^2(t) \left(dx^2+dy^2+dz^2 \right)
 \end{equation}
 where a(t) and b(t) are given by the timt-time and space-space components of the tensor field $ h^{\mu \nu}$ 
. The evolution equation for a(t) in massive gravity has the form 
\begin{equation}
3 \left ({\dot{a(t)} \over a(t)} \right)^2+{3 \alpha^2 \over {8(\zeta+2)}}
[y^3-(1-4\zeta){1 \over y}+{2\zeta \over {a^2}}(y^2-3) ]={{8 \pi G\rho_0}\over a^{3(\omega+1)}}
\end{equation}
where $\alpha^2=4k_1$, $\beta^2=-2k \left(k_1+4k_2 \right)/(k_1+k_2)$,  $\zeta = \beta^2/\alpha^2$. The perfect fluid form of $T_{\mu \nu}$ is assumed and  the equation of state is taken in the simple form, $p= \omega \rho$ with $\omega$ constant.
In contrast to GR the function b(t) is not arbitrary and gets determined through a(t) itself ; 
there is an algebraic relation between a(t) and b(t) \cite{babak1}. This relation acquires a simple form for $\zeta =1/4$
\begin{equation}
y \equiv {a \over b}={(-1+\sqrt{7+9a^4}) \over {3a^2}}
\end{equation}

\section{Phase Space Analysis}
In this section we  investigate the phase space behavior of cosmological evolution \cite{belinsky} in finite range gravity for both the gravitons
behaving like tachyons, i.e, $\alpha^2, \beta^2 < 0$. For simplicity, we will  be working with $\zeta=1/4$. The evolution equation 
( 4 ) in this case for $ a>>1$ and $ \omega=0$, where interesting things are expected to happen, simplifies to
\begin{equation}
a'^2(\tau)={a^{-1} \over 3}+\Lambda_ma^2+{7 \over 6}\Lambda_m a^{-2}+{7 \over 18}\Lambda_m a^{-4}-\Lambda_m
\end{equation}
and for $a<<1$
\begin{equation}
a'^2\simeq {\Lambda_m \over 3} a^{-4}+{ a^{-1} \over 3}
\end{equation}
where $\Lambda_m=\left({{\left |\alpha \right |^2} \over {8(\zeta +2)}}\right){1 \over {8 \pi G \rho_0}}$, and prime denotes the derivative with respect to $\tau=\sqrt{(8\pi G \rho_0)} t$. In order to obtain the expression ( 6 ) we have used the following series expansion for $y ( a>>1 )$
$$y=1+{7 \over 9} a^{-4}+ higher~ order~ terms$$
Before proceeding further, a comment about Eq.( 6 ) is in order. From the point of view of GR ,Equation ( 6 ) can be thought as  an effective Friedmann equation where the  
first term on the right hand side of this equation is the usual term contributed by the matter density and the rest of the terms containing $\Lambda_m$ originate from
the Mass tensor appearing in the finite range gravity theory.  
 The second term behaves like cosmological constant and is responsible for 
drawing the late time accelerated expansion in Babak-Grishchuk theory. It is remarkable that such a term automatically appears
at late stages of evolution in massive gravity. The terms containing $ \Lambda_m$ smoothly vanish in the mass-zero limit leaving behind the usual Freedmann equation. It is surprising that an arbitrarily small mass of gravitons can change the
character of evolution, turning deceleration into acceleration.The third term in Eq. 6 mimics radiation whereas the last term
looks like curvature term. It is interesting to notice that Eq. 6  looks formally similar to the effective Friedmann equation in brane-world
cosmology \cite{langlois}. It should however be kept in mind that unlike the FRW cosmology where the numerical value of the scale factor does not carry any physical significance and  the scale factor  can be made larger
or smaller than one at any given epoch,  such a choice no longer exists in massive gravity. As already mentioned, the function b(t)
is not arbitrary in  Babak-Grishchuk theory, it gets determined through  a(t) via Eq. 4 which clearly shows that the numerical value of scale factor is  indeed important.
Eq. 6 is valid for large values of "a" and in this regime out of all the terms containing $\Lambda_m$, the second term is most dominant which is to compete with the first term allowing ultimately the transition to acceleration. 
 It is further important to mention, as pointed out by Babak and Grishchuk\cite{ babak1}, that for no values of parameters, $\dot{a}=0$ in the present case ( however, for $\alpha^2, \beta^2 >0$, the scale factor goes through regular maximum and minimum). 
At early times, the evolution of universe is described by Eq. 7. The first term on the right hand side (RHS) of this equation
dominates at early epochs and mimics stiff equation of state.\par
It would be suggestive to recast the expression ( 6 ) to look like Newton equation in one dimension\cite{sahni}
\begin{equation}
a^{''}(\tau)=-{d \over da} V(a)
\end{equation}
with the first integral of motion
\begin{equation}
E={a'^2 \over 2}+V(a)=0
\end{equation}
where $V(a)$ is given by the expression,
\begin{equation}
V(a)=-{a^{-1} \over 6}-{\Lambda_m \over 2} { a^2 }- {7 \over 12} \Lambda_m a^{-2}-{ 7 \over 36}\Lambda_m a^{-4}+{\Lambda_m \over 2}
\end{equation}

 The potential V(a) is plotted in Fig.1. The system, in the present case, has sufficient kinetic energy to surmount the
potential barrier and, consequently, the scale factor passes through a point of inflection at which the deceleration changes into acceleration.
\begin{figure}
\resizebox{3.0in}{!}{\includegraphics{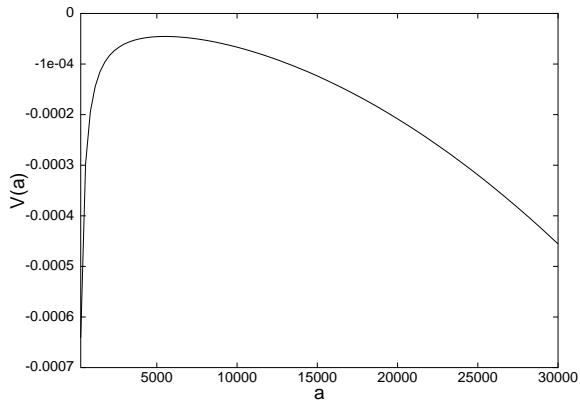}}
\caption{
Plot of the potential V(a) given by Eq.10 for $ \Lambda_m=10^{-12}$. This is a generic form of potential which in case of specially flat universe allows deceleration to change into acceleration.}

\end{figure}
In order to analyze the dynamics of the system we recast the equation ( 8 ) in the following form
\begin{equation}
v'=-uv
\end{equation}
\begin{equation}
u'=\Lambda_m-{7 \over 6}\Lambda_m v^4-{7 \over 9}\Lambda_m v^6-{v^3 \over 6}-u^2
\end{equation}
where
$$ v={1 \over a},~~~~~ u={a' \over a}$$. The constraint equation ( 9 ) takes the form

\begin{equation}
{u^2 \over 2}={\Lambda_m \over 2}-{\Lambda_m \over 2}v^2+{7 \over 12}\Lambda_m v^4+{7 \over 36}\Lambda_m v^6+{v^3 \over 6}
\end{equation}
\begin{figure}
\resizebox{3.0in}{!}{\includegraphics{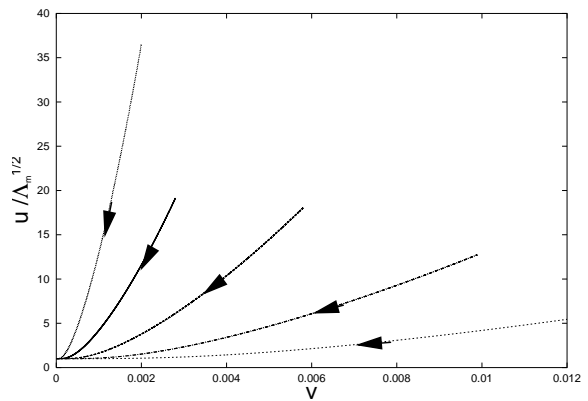}}
\caption{
 Phase portrait of cosmological evolution in massive gravity for $ \alpha^2, \beta^2 <0$ , i.e., for both the gravitons in the theory behaving like tachyons.Trajectories, corresponding to different values of parameter $\Lambda_m$, starting anywhere in the phase space, consistent with the constraint given by Eq. (13), end up at the stable critical point (0,1).}
\end{figure}
To find out the critical points we set
right-hand side (RHS) of Eqs. (11) and (12) to zero and we obtain
 two fixed points
$$u_{cr}=0,~~~~~   v_{cr}=v_0            $$
and
$$v_{cr}=0  ,~~~~~ u_{cr}=+\Lambda^{1/2}$$
where $v_0$ is a solution of the algebraic equation
\begin{equation}
{7 \over 9}\Lambda_m v_0^6+{7 \over 6}\Lambda_m v_0^4+{v_0^3 \over 6}-\Lambda_m=0
\end{equation}
$v_{cr}=-(\Lambda_m)^{1/2}$ is not relevant here.
At the first critical point, ${a'(\tau)}=0$ and this is not possible in the 
present  case i .e, for $\alpha^2, \beta^2<0$.
Indeed, omitting terms of higher order than $ v^3$ in equation ( 14 ), one gets the approximate
value of $ v_0 \simeq ( 6 \Lambda_m )^{1/3}$. This critical point violates the constraint equation ( 13 )  and therefore  should be discarded.
 The second critical point which satisfies the constraint equation ( 13 ) is very interesting, 
$u_{cr}=+(\Lambda_m)^{1/2}$ means that 
$${a'(\tau)}/a(\tau)=\Lambda_m^{1/2}~~~~or~~~~~~~~~    a(\tau) \propto e^{\Lambda_m^{1/2} \tau}   $$
and $v_{cr}=0$ means that the exponential behavior is  approached asymptotically for large values of $ a(\tau)$. Thus, this scenario describes a rapidly accelerating universe eternally at late times and like any other generic model
of quintessence, the finite-range theory of gravity will also be faced with the problem of future event horizon\cite{fischler}. In order to check  the stability of the critical point we perturb about
fixed point $( 0, +\Lambda^{1/2} )$
\begin{eqnarray*}
v &=& v_{cr}+\delta v\\
u &=& u_{cr}+\delta u
\end{eqnarray*}
Plugging these in ( 11 ) and ( 12 )  and keeping only linear terms we get
\begin{equation}
{\delta v}' = -u_{cr} \delta v   
\end{equation}
\begin{equation}          
{\delta u}' = -2u_{cr} {\delta u }
\end{equation}
with the solutions
$$ \delta v \propto e^{-\Lambda^{1/2} \tau}~~~~~~and~~~~~~\delta u \propto e^{-2\Lambda^{1/2} \tau} $$
which clearly demonstrate the stability of the critical point under small perturbations. As the system evolves, the point moves in the (u, v) plane along
the curve given by the constraint equation ( 13 ) on which lies the critical point.
 Stability guarantees that the motion will take place towards the critical pont. We have evolved the system numerically for 
different values of the parameter $\Lambda_m$, we find that trajectories starting
anywhere in the phase space, consistent with the constraint, end up at the stable fixed point as shown in Fig. 2. Therefore  the accelerated expansion is a late time
attractor of finite range gravity.\par

To summarize, we have considered the cosmological evolution in the massive gravity theory proposed by Babak and Grishchuk. We have
investigated the phase space behavior of the system in case both the massive gravitons in the theory are tachyons. We find that
late time acceleration is an attractor of the system.We therefore conclude that the late time accelerated expansion is a generic solution of Babak-Grishchuk theory which does not require any speculative form of matter to achieve it. It would be interesting to confront
the predictions of Babak-Grishchuk theory with observations and investigate the constraints imposed by supernova results 
 on the massive gravity.
\begin{acknowledgements}
I am thankful to L. P. Grishchuk, Stanislav Babak, V. Sahni, T. Padmanabhan, T. Qureshi and D. V. Ahluwalia for 
useful discussions. I am also thankful to A. S. Gupta for bringing the paper of Babak and Grishchuk to my attention.
\end{acknowledgements}

\end{document}